\documentclass[
aip,
reprint,
amsmath,
amssymb]{revtex4-1}
\usepackage{graphicx}
\usepackage{hyperref}
\hypersetup{
	colorlinks=true,
	linkcolor=blue,
	filecolor=blue,      
	urlcolor=blue,
	citecolor=blue,
}
\usepackage[usenames,dvipsnames]{color}
\usepackage{epsfig,amssymb,amsmath,latexsym}

\newcommand{\UP}{\uparrow}
\newcommand{\DN}{\downarrow}
\newcommand{\mb}{\mathbf}
\usepackage{subfigure}
\usepackage{sidecap,tikz}
\usepackage[normalem]{ulem}
\DeclareRobustCommand{\orcidicon}{\hspace{-1.0mm}
	\begin{tikzpicture}
	\draw[lime, fill=lime] (0.0,0.0) 
	circle [radius=0.15] 
	node[white] {{\fontfamily{qag}\selectfont \tiny \,ID}};
	\draw[white, fill=white] (-0.0525,0.095) 
	circle [radius=0.007];
	\end{tikzpicture}
	\hspace{-3.0mm}
}
\foreach \x in {A, ..., Z}{\expandafter\xdef\csname orcid\x\endcsname{\noexpand\href{https://orcid.org/\csname orcidauthor\x\endcsname}
		{\noexpand\orcidicon}}
}

\begin{document}

\preprint{AIP/123-QED}

\title[]{Multiple Topological Phase Transitions Unveiling Gapless Topological Superconductivity in Magnet/Unconventional Superconductor Hybrid Platform}

\author{Minakshi Subhadarshini}
\altaffiliation{These authors contributed equally to this work.}
\affiliation{Institute of Physics, Sachivalaya Marg, Bhubaneswar-751005, India,} 
\affiliation{Homi Bhabha National Institute, Training School Complex, Anushakti Nagar, Mumbai 400094, India}
\author{Amartya Pal\orcidB{}}%
\altaffiliation{These authors contributed equally to this work.}
\affiliation{Institute of Physics, Sachivalaya Marg, Bhubaneswar-751005, India,} 
\affiliation{Homi Bhabha National Institute, Training School Complex, Anushakti Nagar, Mumbai 400094, India}

\author{Pritam Chatterjee\orcidC{}}
\affiliation{Institute of Physics, Sachivalaya Marg, Bhubaneswar-751005, India,} 
\affiliation{Homi Bhabha National Institute, Training School Complex, Anushakti Nagar, Mumbai 400094, India}

\author{Arijit Saha\orcidD{}}
\email{arijit@iopb.res.in}
\affiliation{Institute of Physics, Sachivalaya Marg, Bhubaneswar-751005, India,} 
\affiliation{Homi Bhabha National Institute, Training School Complex, Anushakti Nagar, Mumbai 400094, India}

\date{\today}

\begin{abstract}
We propose a theoretical framework for generating gapless topological superconductivity (GTSC) hosting Majorana flat edge modes (MFEMs) in the presence of a two-dimensional (2D) array of 
magnetic adatoms with noncollinear spin texture deposited on top of a unconventional superconductor. Our observations reveal two distinct topological phase transitions within the emergent Shiba band depending on the exchange coupling strength ($J$) between magnetic adatom spins and superconducting electrons: the first one designates transition from gapless non-topological to gapless topological phase at lower $J$, while the second one denotes transition from gapless topological to a trivial gapped superconducting phase at higher $J$. The gapless topological superconducting phase survives at intermediate values of $J$, hosting MFEMs. Further, we investigate the nature of the bulk effective pairings which indicate that GTSC appears due to the interplay between pseudo ``$s$-wave" and pseudo ``$p_{x}+p_y$" types of pairing. Consequently, our study opens a promising avenue for the experimental realization of GTSC in 2D
Shiba lattice based on $d$-wave superconductors as a high-temperature platform.     
\end{abstract}

\maketitle

Topological superconductivity (TSC) hosting Majorana zero modes (MZMs) has emerged as one of the fascinating research areas in the modern quantum condensed matter physics community, due to its potential application in topological quantum computation and memory storage applications~\cite{Kitaev2001,Ivanov2001,SDSarma2008,Kitaev2009,Alicea_2012,Leijnse_2012,beenakker2013search}. There are several theoretical proposals that exist in the literature, which introduce the generation of isolated MZMs and their possible experimental feasibility~\cite{Kitaev2001,Kitaev2009,Read2000,Fu2008,Oreg2010,Lutchyn2010,Kanasugi2019}. Few experimental setups have been fabricated based on these theoretical proposals, but they still pose significant challenges~\cite{Mourik2012,Das2012,Rokhinson2012,Finck2013,Albrecht2016,Deng2016,Schumann2020,Kaveh2019}. In recent times, one of the most promising ways to generate TSC hosting MZMs relies on magnetic adatoms~
\cite{Felix2013,AliYazdani2013,DanielLoss2013,PascalSimon2013,MFranz2013,Eugene2013,Felix2014,TeemuOjanen2014,MFranz2014,Rajiv2015,Sarma2015,Hoffman2016,Jens2016,Tewari2016,PascalSimon2017,Simon2017,Theiler2019,Cristian2019,Mashkoori2019,Menard2019,Pradhan2020,Teixeira2020,Alexander2020,Perrin2021,Nicholas2020,Chatterjee2023,chatterjee2023b,Mondal2023,Jelena2016,Balatsky22016,Eigler1997,Yazdani1999,Yazdani2015,Wiesendanger2021,Beck2021,Wang2021,Schneider2022,Richard2022,Wiesendanger2022,Yacoby2023,Soldini2023}. According to this proposal, an array of magnetic adatoms with classical spins is fabricated on the surface of a bulk $s$-wave superconductor. This heterostructure can give rise to isolated MZMs at the end of the chain due to the scattering between the classical spin of magnetic adatoms and the spin of the superconducting electrons. Another interesting aspect of this setup  
is the formation of an emergent band within the superconducting gap in the presence of magnetic impurities. This is called the Yu-Shiba-Rusinov (YSR) band or simply the Shiba band~\cite{Felix2013,AliYazdani2013,Shiba1968}. Experimentally, researchers have also observed the existence of the Shiba band~\cite{Eigler1997,Yazdani1999,Yazdani2015,Wiesendanger2021,Beck2021,Wang2021,Schneider2022,Richard2022,Wiesendanger2022,Yacoby2023,Soldini2023}, which plays the pivotal role during the topological phase transition. The sign change of the Shiba minigap indicates topological superconducting phase transition hosting MZMs~\cite{Felix2013,AliYazdani2013,Yazdani2015}.

In recent times, there has been a growing interest in the generation of MFEMs over MZMs for the memory storage applications~\cite{Wang2017,Zhang2019,Wong2013,Deng2014,Nakosai2013,Bena2015,chatterjee2023c}, which is a direct consequence of the 2D Kitaev model~\cite{Wang2017,Zhang2019}. 
In literature, the generation of MZMs involving Shiba states have been reported in several theoretical and experimental works based on unconventional superconductors~
\cite{Pientka2014,Kreisel2021,Neupert2016,Crawford2020,Ghazaryan2022,Chatzopoulos2021}. However, only a handful of recent articles have explored the engineering of MFEMs via magnet-superconductor heterostructures~\cite{Bena2015,chatterjee2023c}. Moreover, these proposals are based on conventional $s$-wave superconductors~\cite{Bena2015,chatterjee2023c}. 
Till date, no article is available in the literature, 
where the generation of MFEMs has been proposed resulting from the coexistence of magnet and unconventional superconductor hybrid system.

\begin{figure}[h]
	\centering
	\subfigure{\includegraphics[width=0.43\textwidth]{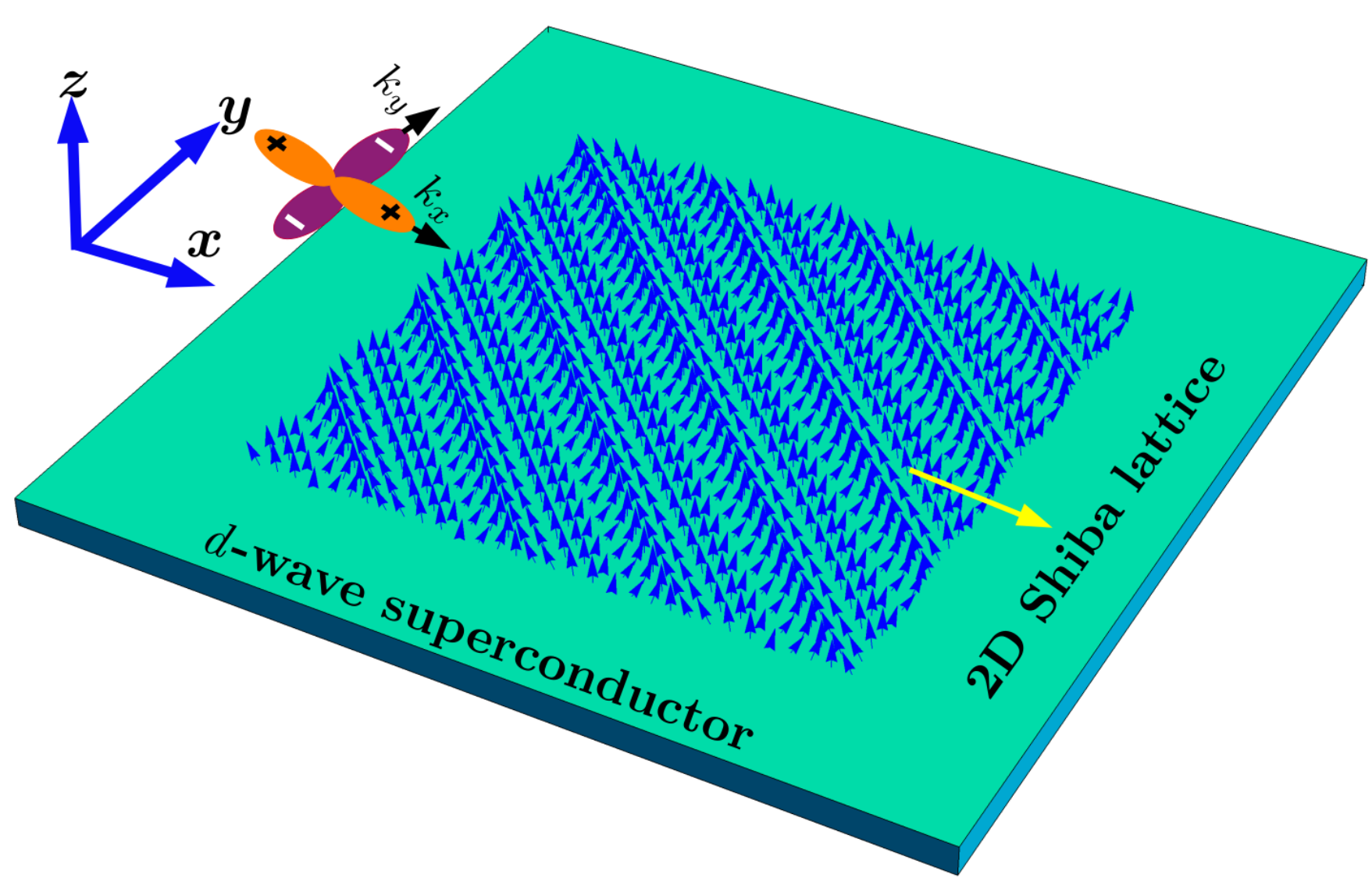}}
	\caption{Schematic diagram of our hybrid setup comprising of a 2D Shiba lattice with noncolinear magnetic texture (blue arrows), placed on top of a unconventional $d$-wave SC (green) with 
	the $d$-wave nature of the pairing gap depicted in the $x$-$y$ plane (top left).}
	\label{Fig1}
\end{figure}

In this article, we put forward a theoretical proposal for the emergence of the GTSC phase hosting MFEMs in the presence of a 2D array of noncollinear magnetic texture deposited on top of a 
$d$-wave superconductor (see Fig.~\ref{Fig1} for a schematic view). Presence of exchange field inside a $d$-wave superconductor (SC) generates inflated topological Fermi surfaces~\cite{Yang1998,Setty2020,Setty2020NatComm,pal2023}. However, in this article, we primarily focus on the generation of MFEMs using noncolinear magnetism. Interestingly, we obtain two distinct topological phase transitions within the emergent 2D Shiba band, depending on the exchange coupling strength ($J$) between magnetic adatom spins and superconducting electrons: (i)~the first one designates gapless non-topological to gapless topological phase transition at lower $J$, and (ii) the second one from gapless topological to a trivial gapped superconducting phase at higher $J$. The gapless topological phase resides at intermediate values of $J$, resulting in GTSC hosting MFEMs. Moreover, we compute the bulk effective pairings using a duality transformation of the low-energy continuum Hamiltonian. It suggests that the GTSC can be stabilized due to the interplay between pseudo ``$s$-wave" and pseudo ``$p_{x}+p_{y}$" types of pairings.


To begin with, we architect a toy model based on a 2D array of magnetic adatoms with noncollinear spin texture deposited on a unconventional $d$-wave superconductor as schematically shown in Fig.\,\ref{Fig1}. The Bogoliubov de Gennes (BdG) Hamiltonian in the continuum limit can be written as, $\mathcal{H} = \int d \mathbf{r} \, \Psi^{\dagger} (\mathbf{r})  \mathcal{H}_{\rm{BdG}}  \Psi(\mathbf{r})$ where, $\Psi(\mathbf{r})= [c_{\mathbf{r},\uparrow}, c_{\mathbf{r},\downarrow}, c_{\mathbf{r},\downarrow}^{\dagger},-c_{\mathbf{r},\uparrow} ^{\dagger}]^{T}$ is the Nambu spinor with $c_{r,\sigma}$~
($c^\dagger_{r,\sigma}$) being the electron annhilation (creation) operator with spin $\sigma(=\uparrow,\downarrow)$ at position $\mathbf{r}=(x,y)$. The BdG Hamiltonian in the first quantized form 
can be written as,
\begin{eqnarray}
\mathcal{H}_{\rm{BdG}}\!=\! & -&\frac{\hbar^2}{2m} \boldsymbol{\nabla}^2\tau_z + J  \mathbf{S}(\mathbf{r}).\boldsymbol{\sigma} - \mu \tau_z\!  +\!\Delta_d(\mathbf{r})\tau_x\ , 
\end{eqnarray}
Here, the Pauli matrices, $\tau$ and $\sigma$, act on particle hole and spin degrees of freedom respectively. $J$ denotes the exchange coupling strength between magnetic impurites and superconducting electrons. Spin of the magnetic impurities is assumed to be classical and represented by an unit vector as $\mathbf{S(r)}= (\sin \theta_r\cos \phi_r,\sin \theta_r \sin\phi_r,\cos\theta_r)$, where $\theta_r,
\phi_r$ being the polar and azimuthal angle at position $\mathbf{r}$. The symbols $\mu$ and $\Delta_d(\mb{r}) (=\Delta_0 (\partial_x^2-\partial_y^2))$ denotes the chemical potential and $d$-wave pairing amplitude of the superconductor respectively. For simplicity, we assume $\hbar=m=1$ throughout the paper. 

We perform two successive unitary transformations, $U_1=e^{-i(\phi_r/2-\pi/4)\sigma_z}$ and $U_2=e^{-i(\theta_r/2-\pi/4)\sigma_x}$ respectively, to obtain a translationally invariant low energy effective Hamiltonian as, $\mathcal{H}_{\rm{eff}}(\mb{r})=U_2^\dagger\, U_1^\dagger\, \mathcal{H}_{\rm{BdG}}(\mb{r})\,U_1\,U_2$~\cite{Hess2023,chatterjee2023c,chatterjee2023b}. Further, we choose, 
$\theta_r=(g_xx+g_yy)$ and $\phi_r=0$ throughout the calculation. We also consider a homogeneous spin spiral (SS) by choosing $g_x=g_y=g$ for simplicity. The symbol $g$ represents the pitch vector 
of the SS. The low energy effective Hamiltonian in the momentum-space can be written as,

\begin{equation}
\mathcal{H}_{\rm{eff}}(\mb{k})=\tilde{\xi}_\mb{k}\tau_z \!+\!\frac{g}{4}(k_x\!+\!k_y) \tau_z\sigma_x 
+ \Delta_{\rm{eff}}\tau_x \!+\! J\sigma_y\ , \label{Eq.Effective_Ham}
\end{equation}
where, $\tilde{\xi}_\mb{k}=\frac{1}{2} (\mb{k}^2 \!+\! \frac{g^2}{2})-\!\mu$ and $\Delta_{\rm{eff}}\!\!=\!\!\Delta_0[(k_x^2-k_y^2) + \frac{g}{2}(k_x\!-\!k_y)\sigma_x]$. The term $\Delta_{\rm{eff}}$ is comprised of the parent $d$-wave pairing amplitude and in addition, an emergent term $(p_x \!- \!p_y)$ appears due to the presence of the 2D SS. The second term in Eq.~\eqref{Eq.Effective_Ham} acts as an 
effective 2D spin orbit coupling (SOC) while the last term in Eq.~\eqref{Eq.Effective_Ham} denotes an effective in plane Zeeman field along $y$-direction. 

\begin{figure}[h!]
	\centering
	\subfigure{\includegraphics[width=0.5\textwidth]{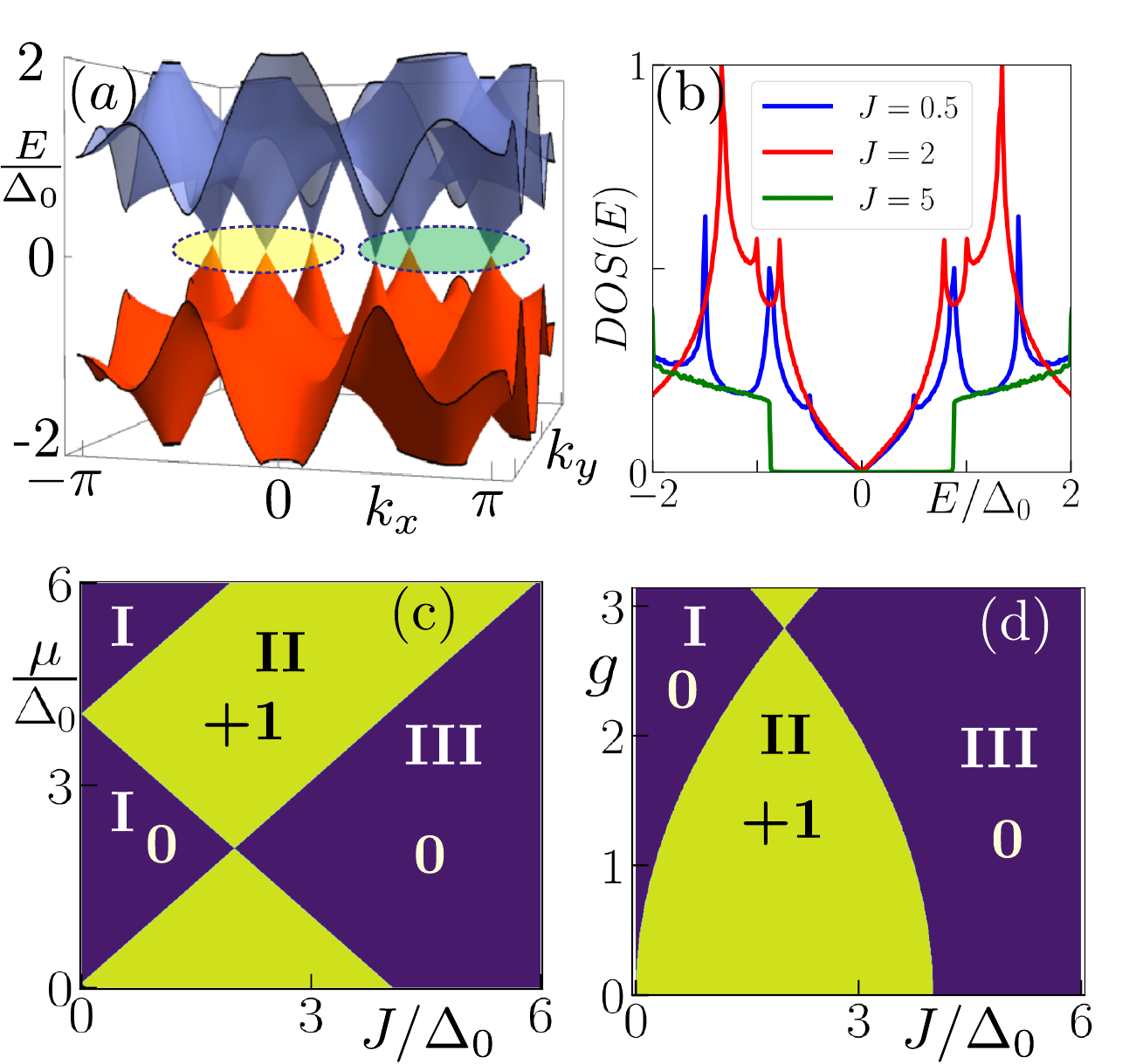}}
	\caption{Panel (a) displays the bulk band spectrum within the topological regime ($J=2$), featuring six band touching points. In panel (b), the bulk DOS is depicted for three phases: non-topological 		gapless ($J=0.5$), topological gapless ($J=2$), and gapped non-topological ($J=5$). We choose $g=2$ for panels(a)-(b). The invariant, $\nu$, is demonstrated in the $\mu/\Delta_0-J/\Delta_0$ 
	plane choosing $g=0.5$ in panel (c) and in the $g/\Delta_0-J/\Delta_0$ plane in panel (d) with $\mu=4$.}
	\label{Fig2}
\end{figure}

Furthermore, we compute the bulk topological invariant in order to charectarize the gapless topological superconducting phase starting from the lattice regularized version of the Hamiltonian mentioned
in Eq.~\eqref{Eq.Effective_Ham}. We replace $k_{x,y}\rightarrow \sin k_{x,y}$ and $k_{x,y}^2\rightarrow2(1-\cos k_{x,y})$. The $\mathcal{Z}_2$ topological invariant denoted by $\nu$, is defined as,~\cite{Bena2015,chatterjee2023c}
\begin{equation}
(-1)^\nu = \mathrm{sgn}  \prod_{i=1}^{4} \det 
\begin{pmatrix}
\xi_{\mb{k}}(\Gamma_i) - J & \Delta_d(\Gamma_i)\\
\Delta_d(\Gamma_i) & -\xi_{\mb{k}}(\Gamma_i)-J
\end{pmatrix}\ ,
\end{equation}
where, $\nu \!=\!\!0$ and $\nu\!\!=\!\!1$ represent the topologically trivial and nontrivial regime respectively. The symbol $\boldsymbol{\Gamma}=[(0,0),(0,\pi),(\pi,0),(\pi,\pi)]$ denotes the four high symmetry points of the square brillouin zone. In Fig.~\ref{Fig2}(c) and Fig.~\ref{Fig2}(d), we illustrate the topological invariant in $\mu-J$ and $g-J$ plane respectively. Interestingly, we observe two topological phase transitions depending on the exchange coupling strength $J$ for fixed values of $\mu$ and $g$. We obtain two types of topologically trivial phases with $\nu=0$ (labelled as I and III 
in Fig.~\ref{Fig2}) separated by a topologically non-trivial phase with $\nu=1$ (indicated as II in Fig.~\ref{Fig2}). To further investigate these different phases, we compute the bulk density of states (DOS), 
${\rm{DOS~(E)}}=(1/\pi)\sum_\mb{k} \delta(E-E_\mb{k})$ which is depicted in Fig.~\ref{Fig2}(b) for three distinct values of $J$, representing three phases I, II and III. The latter is denoted by blue, red and green color line respectively. Thus, the system undergoes two phase transitions from (i) gapless non-topological to gapless topological phase and (ii) gapless topological to gapped non-topological phase
within the emergent Shiba band [see Figs.~\ref{Fig2}(c)-(d)]. We identify the GTSC with six band touching points associated with semimetallic DOS in the topological regime (phase II) from bulk band structure and DOS as shown in Figs.~\ref{Fig2}(a)-(b). On the other hand, two non-topological phases exhibit both gapped and semimetallic DOS tunable by exchange coupling strength $J$ [see Fig.~\ref{Fig2}((b)]. 

To examine the nature of bulk effective pairing we perform a unitary transformation,~\cite{Sato2009,Sato2010,chatterjee2023b} 
in order to obtain a dual Hamiltonian of Eq.~\eqref{Eq.Effective_Ham}, $\tilde{\mathcal{H}}_{\rm{D}}  = \tilde{U}^\dagger \mathcal{H}_{\rm{eff}}\tilde{U}$ as, 
\begin{equation}
\tilde{U}\!\!=\!\!\frac{1}{\sqrt{2}}\begin{pmatrix}
1 & -1\\
1 & 1 \\
\end{pmatrix}\sigma_0;\,
\tilde{\mathcal{H}}_{\rm{D}} \!\!=\!\!\begin{pmatrix}
\!\tilde{\xi}_{\mb{k}}^{\rm{D}}  \sigma_z + J\sigma_y & \!\!\!\!\tilde{\Delta}_{\rm{D}}  \\
\!\!\!\!\tilde{\Delta}_{\rm{D}}  &\!\!\!\! -\tilde{\xi}_{\mb{k}}^{\rm{D}}  \sigma_z +J\sigma_y
\end{pmatrix}\ ,
\label{Eq4} 
\end{equation}

where, $\tilde{\xi}_{\mb{k}}^{\rm{D}} =\frac{g}{2}\left(k_x - k_y \right)$ and $\tilde{\Delta}_{\rm{D}}=\tilde{\Delta}_s\sigma_0 +\tilde{\Delta}_p \sigma_x =(2-\mu + g^2/4)\sigma_0 + (g/4)(k_x + k_y) \sigma_x$ are called dual kinetic energy and dual SC like gap of the transformed Hamiltonian. Here we neglect $O(k^2)$ terms at the low energy limit. The dual spinor takes the form in pseudo Nambu basis as, 
$\tilde{\Psi}(\mathbf{k})= [ \tilde{c}_{\mb{k},+}, \tilde{c}_{\mb{k},-}, \tilde{c}_{\mb{k},-}^\dagger,-\tilde{c}_{\mb{k},+} ^{\dagger}]^{T}$, where, $\tilde{c}_{\mb{k},+(-)} = (c_{\mb{k},\UP(\DN)} \,+ \!(-)\, c^\dagger_{-\mb{k},\DN(\UP)})/\sqrt{2}$ mimics a system with pseudo-spin degree of freedom. Therefore, the effective pairing/dual gap ($\tilde{\Delta}_{\rm{D}}$) turns out to be the combination of a pseudo $s$-wave ($\tilde{\Delta}_s$) and a pseudo $(p_x+p_y)$ ($\tilde{\Delta}_p$) type pairings in the topological phase (II). Hence, the emergence of the pseudo $(p_x + p_y)$ type pairing is the direct consequence of the nonlinear magnetic texture that stabilizes the gapless topological phase II hosting MFEMs~\cite{Wang2017,Zhang2019,Nakosai2013,Bena2015,chatterjee2023c} which we discuss 
in detail in the latter text.

\begin{figure}[h!]
	\centering
	\subfigure{\includegraphics[width=0.5\textwidth]{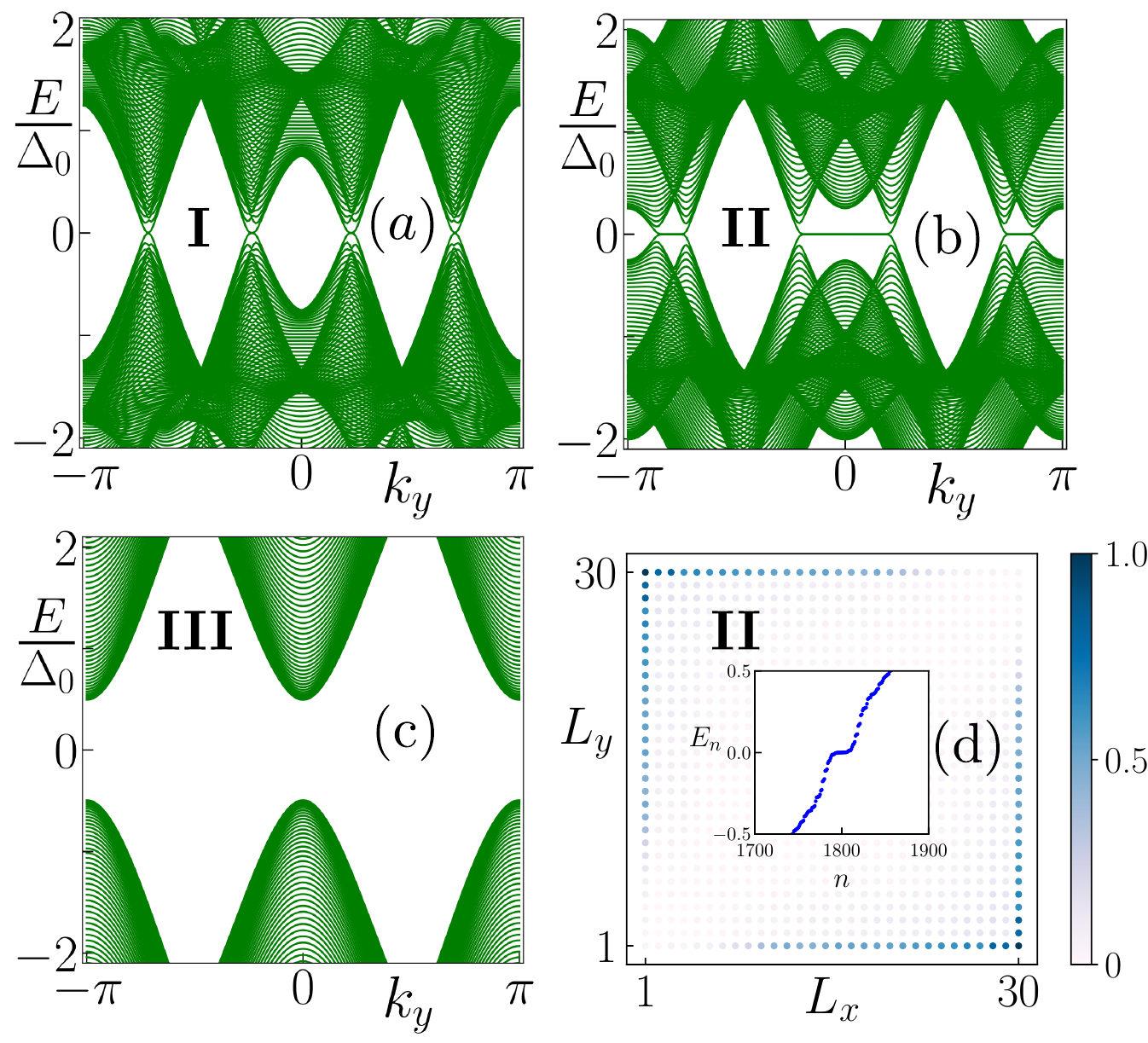}}
	\caption{
		Panels (a)-(c) depict the energy eigenvalue spectrum considering the ribbon geometry, revealing the presence of multiple zero-energy flat edge modes in the topological regime [(b) $J=2$]. 
		This stands in sharp contrast to the absence of such modes in the trivial region [(a) $J=0.5$, and (c) $J=5$]. Panel (d) displays the normalized site resolved LDOS at $E=0$
		corresponding to the MEFM, computed within a $30\times 30$ square lattice. In all cases, we choose the model parameters as $\Delta_0=1$, $t=1$, $g=3$ and $\mu=4$.}
	\label{Fig:Ribbon_Geometry}
\end{figure}

To investigate the boundary of our 2D system, we analyze our results based on finite geometry calculations performed via the lattice-regularized version of the Hamiltonian in Eq.~\eqref{Eq.Effective_Ham}. In Figs.~\ref{Fig:Ribbon_Geometry}(a)-(c), we illustrate the eigenvalue spectrum as a function of momentum $k_y$, considering open boundary conditions (OBC) along the $x$ direction and periodic boundary conditions (PBC) along the $y$ direction. Therefore, momentum along the $y$ direction ($k_y$) is a good quantum number. Here, Fig.~\ref{Fig:Ribbon_Geometry}(a), Fig.~\ref{Fig:Ribbon_Geometry}(b), and Fig.~\ref{Fig:Ribbon_Geometry}(c) correspond to the three phases I, II, and III, respectively, as depicted in Fig.~\ref{Fig2}. Interestingly, the topological phase (phase II) unveils the emergence of three gapless MFEMs between bulk band touching points [see Fig.~\ref{Fig:Ribbon_Geometry}(b)]. This is a consequence of six gapless nodes in the bulk [see Fig.~\ref{Fig2}(a)]. Hence, qualitatively, we can predict that the GTSC anchoring MFEMs can be stabilized due to the presence of pseudo ``$s$-wave" and pseudo ``$p_x+p_y$" type pairing (as discussed before). On the other hand, in the non-topological phases (phase I and III in Fig.~\ref{Fig2}), the system is either gapless or gapped without hosting any types of edge mode as shown in Figs.~\ref{Fig:Ribbon_Geometry}(a) and (c). In Fig.~\ref{Fig:Ribbon_Geometry}(d), we compute the normalized local density of states (LDOS) at $E=0$ in the $L_x-L_y$ plane associated with phase II. We consider OBC along both the $x$ and $y$ directions, employing the formula $N_i(E)=\sum_{n} |\phi_n(i)|^2\delta(E-E_n)$, where $\phi_n(i)$ is the eigenstate of the Hamiltonian at site $i$. The symbol $n$ is the eigenvalue index with eigenvalue $E_n$. It is evident from the LDOS distribution that zero-energy eigenstates are maximally localized at the edges (MFEMs), which is also reflected in the $E_n$ vs $n$ behavior via the inset of the same figure. Note that, the location of the edge modes are not constrained to be appeared at the negative diagonal of the 2D domain. Edge states can also appear in the positive diagonal of the system by changing the spin texture configuration from $\theta_r=(g_xx + g_yy)$ to $\theta_r=(g_xx-g_yy)$ in the effective Hamiltonian. This observation confirms that our model harbors Majorana flat modes localized at the edges, thereby establishing the bulk-boundary correspondence of the system.    

\begin{figure}[h]
	\includegraphics[scale=0.305]{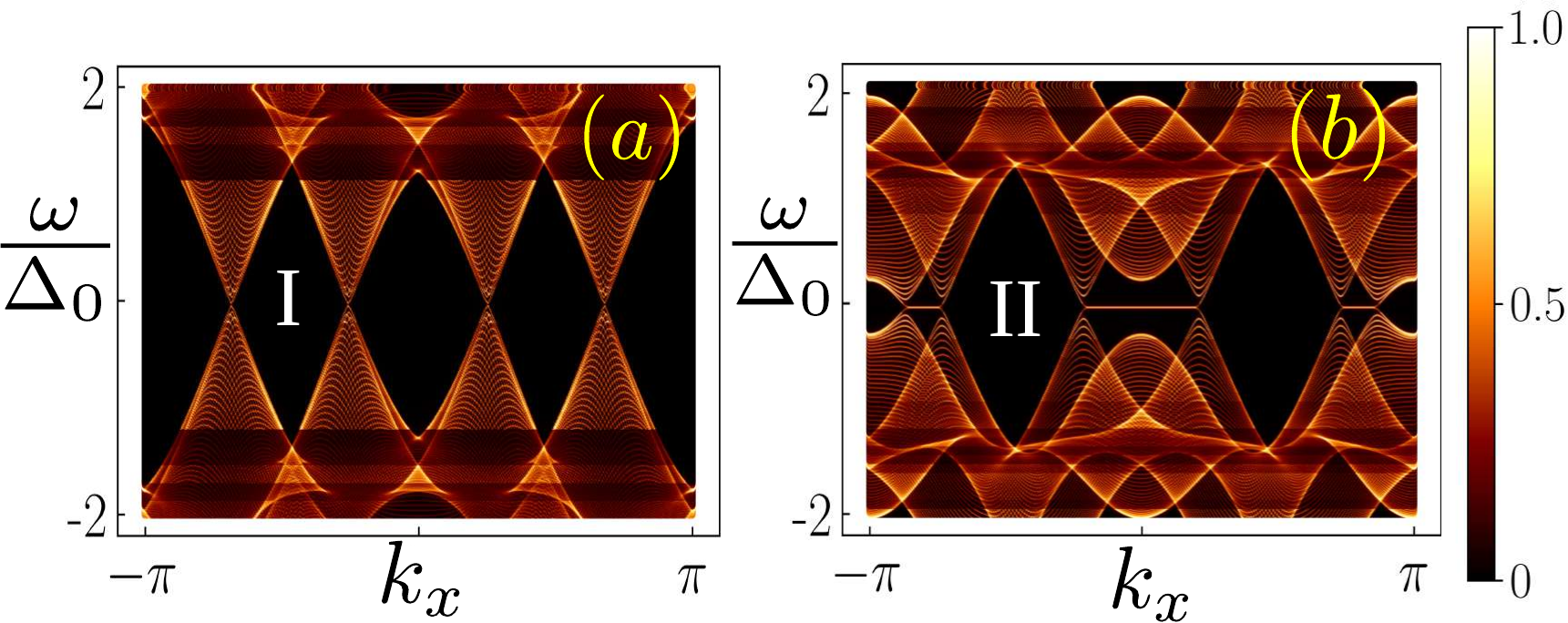}
	\caption{Spectral function, $\mathcal{A}(k,\omega)$ is shown in ($\omega/\Delta_0 - k_x$) plane 
	for ($a$) gapless non-topological phase ($J=0.5$), ($b$) gapless topological phase ($J=2$). 
	We choose the model parameters as $\Delta_0=1,\, t=1, \, g=3 $ and $\mu = 4$.}
	\label{Fig:4 Akw}
\end{figure}

Furthermore, we compute the spectral function, $\mathcal{A}({k_x,\omega})=-\frac{1}{\pi}\mathrm{Im[Tr}\,{\mathcal{G}(k_x,\omega)}]$ with $\mathcal{G}(k_x,\omega)\!\!=\!\![(\omega + i\epsilon)\mathcal{I} - \mathcal{H}(y,k_x)]$ using the lattice regularized Hamiltonian of Eq.~(\ref{Eq.Effective_Ham}) and employing OBC along $y$ direction and PBC along $x$ direction. 
In Fig.~\ref{Fig:4 Akw}, we depict the behavior of $\mathcal{A}(k,\omega)$ in ($\omega/\Delta_0 - k_x$) plane. Here, we discuss the two phases: (i) the gapless non-topological phase and (ii) the gapless topological phase. We observe a clear signature of MFEMs in the topological regime [see Fig.~\ref{Fig:4 Akw}(b)]. On the other hand, MFEMs disappear in the gapless non-topological regime as shown in 
Fig.~\ref{Fig:4 Akw}(a). From practical point of view, it can be possible to measure $\mathcal{A}({k_x,\omega})$ in terms of angle-resolved photoemission spectroscopy (ARPES) to look for the signature 
of MFEMs.


In this article, we systematically investigate the emergence of GTSC phase hosting MFEMs in the presence of a 2D array of a magnetic adatoms characterized by a spatially modulated noncollinear spin texture implemented on an unconventional superconductor (with $d$-wave pairing symmetry) as a high-temperature platform. To elucidate further on the system's topological properties, we derive the low-energy effective Hamiltonian in $k$-space through two successive unitary transformations. Furthermore, we compute the $\mathcal{Z}_2$ topological invariant and bulk DOS to characterize the bulk of the system. Importantly, we find three distinct scenarios: (i) a gapless non-topological phase, (ii) a gapless topological phase hosting multiple MFEMs, and (iii) a gapped non-topological phase (as shown in Fig.~\ref{Fig2}). Finally, we unveil the appearance of multiple MFEMs, maximally localized at the system edges in the topological regime, through ribbon geometry and finite-size lattice simulations. We further compute the spectral function numerically which clearly suggests the appearance of the MFEMs at the boundary of the system. However, note that experimentally detecting the MFEMs via 
$\mathcal{A}(k,\omega)$ can be extremely challenging as bulk states can penetrate substantially due to the gapless nature. One possible way to realize a gapped spectrum for our case is to deposit 
a layer of 2D quantum spin Hall insulator on top of the composite system ($d$-wave superconductor + non-collinear spin texture), following the same route as demonstrated in~\cite{chatterjee2023b}. However, the nature of the bulk topology of such composite system is expected to be different, leading to a second-order topological superconductor, which is beyond the scope of the present manuscript and will be presented elsewhere.

 
The possible experimental realization of our setup involves the placement of a monolayer of magnetic adatoms (such as Fe/Cr/Mn) deposited on top of an iron-based superconducting substrate (such as FeSe, $\beta$-Fe${1.01}$Se, LaFeAs, etc.). Such configuration has attracted significant recent attention in the context of experimental realization of topological superconductivity, hosting MZMs in high-temperature platforms~\cite{Dongfei2018, Fong2008, Medvedev2009, Peng2018, Kamihara2008, Peng2018, Chatzopoulos2021}. Notably, the emergence of in-gap Shiba states, in addition to Majorana bound states, has been observed in FeTe${0.55}$Se$_{0.45}$ superconductors~\cite{Peng2018, Chatzopoulos2021}. Therefore, given the experimental progress in this research field, we believe that our theoretical model proposal for GTSC hosting MFEMs is timely and may be possible to realize in future experiments. However, the exact description of experimental techniques and prediction of candidate materials based on our model Hamiltonian are not the subject matter of our present manuscript.

\begin{acknowledgments}
A.P. acknowledge the SAMKHYA: HPC Facility provided at IOP, Bhubaneswar, for numerical computations. We acknowledge  Department of Atomic Energy (DAE), Govt. of India for providing 
the financial support.	
\end{acknowledgments}

\vspace{0.2cm}

\bibliography{bibfile}

\end{document}